\documentstyle[aps,multicol]{revtex}

\draft

\begin{document}

\title{Universality in self-organized critical slope  models}

\author{Alexei V\'azquez and Oscar Sotolongo-Costa}

\address{Department of Theoretical Physics, Faculty of Physics,
Havana University, Havana 10400, Cuba}

\maketitle

\begin{abstract}

The dynamics of critical slope self-organized critical models is
studied, using a previous mapping into a linear interface
depinning model dragged at one end. The model is solved
obtaining the complete set of scaling exponents. Some results
are supported by previous RG developed for constant force linear
interface depinning models but others, like the linear
dependency of the susceptibility with system size, are intrinsic
of this model which belongs to a different universality class.
The comparison of our results with numerical simulations of
ricepile and vortexpile models reported in the literature
reveals that, as in the constant force case, there are two
universality classes corresponding to random and periodic
pinning.

\end{abstract}


\begin{multicols}{2}

\section{Introduction}

There are many natural phenomena where the dynamics does not
take place smoothly but in the form of bursts or avalanches.
Avalanche dynamics has been observed in earthquakes dynamics
\cite{richter}, granular piles \cite{held,frette1},
superconductor vortex piles \cite{superconductors}, the
Barkhausen effect \cite{barkhausen}, crack propagation
\cite{crack}, and more. The avalanches are characterized by
their size $s$ and duration $T$ which in general follow the
power law distributions $P(s)\sim s^{-\tau_s}$ and $P(T)\sim
T^{-\tau_t}$, respectively.

Bak, Tang, and Wiesenfeld \cite{bak2} introduced the notion of
self-organized criticality (SOC) to explain this kind of
dynamics, which seems to be common to many "different"
phenomena. In their early formulation the critical state of a
SOC system is an attractor of its dynamics and therefore there
is no need of fine-tuning to reach the critical state. This idea
was illustrated with very simple cellular automata, the sandpile
models. In sandpile models an integer
\cite{bak2,kadanoff,manna1,frette,nunes1} or continuous
\cite{zhang,olami,bassler} variable $z_i$ is defined in a
$d$-dimensional lattice. $z_i$ can be the height of grains
columns \cite{bak2,kadanoff,nunes1,christensen}, the stress
accumulated in certain fault \cite{olami}, the vortex density
\cite{bassler}, etc. A critical threshold condition for local
relaxation (toppling) is also given, which can be a critical
height $z_i>z_c$ \cite{bak2,kadanoff,manna1,olami}, critical
slope $m_i=z_i-z_{i+1}>m_c$ \cite{kadanoff,frette,bassler}, or
critical Laplacian $l_i=\sum_{nn}z_i-2dz_i>l_c$ \cite{manna2}
threshold condition. When this condition is fulfilled the site
relaxes transferring grains to its nearest neighbors (nn),
otherwise certain amount of grains is added from an external
field.

Critical height models have been extensively studied in the
literature, either by mean field \cite{MF} and field theories
\cite{vespignani1}, renormalization group (RG)
\cite{diaz-guilera,pietronero}, and numerical simulations
\cite{bak2,kadanoff,manna1,zhang,olami}. However, the study of
critical slope models has been in general limited to numerical
simulations \cite{kadanoff,frette,bassler}, which are not so
extensive as in the critical height variant, while theoretical
approaches are almost absent. There are many experimental
situations, sandpiles and vortexpiles for instance, where the
critical slope threshold condition is more appropriate,
requiring a rigorous treatment which captures the general
features of critical slope models. In this direction Paczuski
and Boettcher \cite{paczuski} provided an starting point. They
mapped a critical slope model into a linear interface depinning
(LID) model where the interface is dragged at one end, showing
the existence of common behavior between these models. The
importance of their observation is that LID models can be solved
using continuous approaches \cite{nattermann,narayan1,narayan2}.
Their work was inconclusive because, up to our knowledge, the
problem of an interface moving on a random environment and
dragged at one end has not been solved. However, they observed
that some scaling exponents, but not all, are identical to those
measured for an interface driven uniformly.

In the present work we solve the problem of an interface moving
on a random environment and dragged at one end. We start with a
LID equation with the appropriate boundary conditions. We then
look for the fluctuations around the average in the stationary
state. In the thermodynamic limit these fluctuations satisfy an
equation similar to that of an interface driven uniformly. The
critical exponents are then computed using previous RG
calculations for the uniformly driven interface
\cite{nattermann,narayan1,narayan2}. The dynamic and roughness
scaling exponents are found identical while other exponents
become different. These results show a very good agreement when
compared with numerical estimates of the scaling exponents
reported in the literature.

\section{The model}

Consider the following one-dimensional critical slope model. An
integer variable $z_i$ ($i=1,2,\ldots,L$) is defined. A site
where the slope $m_i=z_i-z_{i+1}$ is greater than a threshold
$m_c$ is said to be active and relaxes according to the toppling
rule
\begin{eqnarray}
z_i\rightarrow z_i-1,
\nonumber\\
z_{i+1}\rightarrow z_{i+1}+1.
\label{eq:0}
\end{eqnarray}
Grains  are added at site $i=0$ at rate $c$ while the other
boundary $i=L-1$ is open. Usually $c$ is assumed very small in
such a way that a new grain is added only when no site is
active, which is the usual separation of time scales assumed in
numerical simulations.

Paczuski and Boettcher \cite{paczuski} noted that instead of
follow the evolution of the column heights $z_i$ one can follow
the evolution of $h_i(t)$, the total number of toppling events
at site $i$ up to time $t$. The evolution rules for $h_i$ can be
easily obtained from those for $z_i$. If we start with an
initial configuration where $z_i=0$ the number of grains at site
$i$ is obtained adding the number of grains received from the
left neighbor ($h_{i-1}$) and subtracting the number of grains
transferred to the right neighbor ($h_i$), i.e.
\begin{equation}
z_i=h_{i-1}-h_i.
\label{eq:0a}
\end{equation}
Using this expression one may compute the local slope
$m_i=z_i-z_{i+1}$, resulting
\begin{equation}
m_i-m_c=h_{i+1}+h_{i-1}-2h_i-m_c,
\label{eq:1}
\end{equation}
On the other hand, $h_i$ increases by one unit every time the
site $i$ topples, i.e. every time $m_i>m_c$, which can be
written as
\begin{equation}
h_i(t+1)-h_i(t)=\Theta(m_i-m_c),
\label{eq:1a}
\end{equation}
where $\Theta(x)$ is the Heaviside unit step function. 

The evolution rules in eqs. (\ref{eq:1}) and (\ref{eq:1a}) are
just the discretized evolution rules for LID models
\cite{leschhorn}, taking $m_i-m_c$ as the force acting on the
interface. The left term of eq. (\ref{eq:1a}) is a discrete time
derivative while $h_{i+1}+h_{i-1}-2h_i$ is the discrete
Laplacian in one dimension. After coarse-graining one obtains
the continuum equation \cite{leschhorn}
\begin{equation} 
\lambda \partial_t h=\Gamma\nabla^2h-m_c+\eta(x,h). 
\label{eq:2} 
\end{equation}
Here $\lambda$ and $\Gamma$ are coarse-graining parameters,
$\lambda$ can be interpreted as a viscosity coefficient and
$\Gamma$ as a surface tension.

$\eta(x,h)$ is a quenched noise which may have different
origins. For instance, the critical slope may be a random
variable reflecting the randomness in the local rearrangements
of sand grains after each toppling. In a more realistic model
$m_c$ is thus a random variable which changes its value from
site to site after each toppling event. When $h(x,t)$ advances,
i.e. the site $x$ topples, a new random slope is assigned. The
randomness in the critical slope is thus reflected in eq.
(\ref{eq:2}) through the quenched noise $\eta(x,h)$. The
quenched noise $\eta(x,h)$ will be assumed a Gaussian noise with
zero mean and noise correlator
\begin{equation}
\langle \eta(x,h)\eta(x^\prime,t^\prime)\rangle=
\delta^d(x-x^\prime)\Delta(t-t^\prime),
\label{eq:2aa}
\end{equation}
where $\Delta(h)$ is a symmetric function, i.e.
$\Delta(-h)=\Delta(h)$. Depending on the boundary conditions for
$\Delta(h)$ one can distinguish two cases
\cite{nattermann,narayan1,narayan2}. For random disorder
$\Delta(h)$ goes to zero for large $h$ while for periodic
pinning forces $\Delta(h)$ is periodic.

The problem will be completely defined after the initial and
boundary conditions are specified. We are interested in the
stationary solution of eq. (\ref{eq:2}) and therefore the
initial condition is irrelevant. Grains will be added to the
system at constant rate $c$ at site $i=0$. Hence, if we start
with an initial configuration where $z_0=0$ the number of grains
at site $i=0$ is obtained adding the number of grains received
from the external field ($ct$), and subtracting the number of
grains transferred to the right neighbor ($h_i$), i.e.
\begin{equation}
z_i=ct-h_i.
\label{eq:0aa}
\end{equation}
Using this expression one may compute the local slope
$m_i=z_i-z_{i+1}$, resulting
\begin{equation}
m_i-m_c=h_1-2h_0-m_c.
\label{eq:0b}
\end{equation}
This evolution rule together with that in eq. (\ref{eq:1a})
leads to the coarse grained equation
\begin{equation} 
\lambda \partial_t h=\kappa\frac{\partial h}{\partial x}+
ct-h-m_c+\eta(x,h), 
\label{eq:0c} 
\end{equation}
which is the boundary condition at $x=0$. However, in the
stationary state the major contribution to the right hand side
of this equation is given by the term $ct-h$ and, therefore,
this boundary condition can be approximated by the more simple
condition
\begin{equation} 
h(0,t)=ct. 
\label{eq:0d} 
\end{equation}
On the other hand, at $x=L-1$ the boundary is open which implies
that the site $x=L$ will never topples, i.e.
\begin{equation} 
h(L,t)=0. 
\label{eq:0e} 
\end{equation}

Eq. (\ref{eq:2}) can be generalized to a $d$-dimensional model.
To be more specific we consider a $d$-dimensional hyper-cube of
linear size $L$ where grains are added at face $x_{||}=0$ and
the boundary $x_{||}=L$ is open. The equation of motion of
$h(x,t)$ will be given by eq. (\ref{eq:2}), where $\nabla^2$
will be now the $d$-dimensional Laplacian, with the boundary
conditions
\begin{equation}
h|_{x_{||}=0}=ct,\ \ \ \ h|_{x_{||}=L}=0.
\label{eq:3}
\end{equation}
Periodic boundary conditions are assumed in the other
directions. Eqs. (\ref{eq:2}) and (\ref{eq:3}) completely
defines the LID model dragged at one end. For $c\rightarrow0$ we
recover the equation of motion proposed by Paczuski and
Boettcher \cite{paczuski}, which describes the dynamics of
critical slope models in the limit of separation of time scales
and no local dissipation.

\section{Stationary state}

Eq. (\ref{eq:2}) is similar to the equation of motion of LID
models. However, in this case the interface is dragged at one
end instead of being driven by a constant force. The drag at one
end makes the problem asymmetric, which leads to a gradient in
$h(x,t)$. It is thus easier to look for an expansion around this
gradient. With this idea in mind we look for a solution of the
form
\begin{equation}
h(x,t)=h_0(x_{||},t)+y(x,t),
\label{eq:4}
\end{equation}
where $h_0(x,t)$ will be determined imposing the constraint
$y|_{x_{||}=0}=y|_{x_{||}=L}=0$ (symmetric boundary conditions
for $y(x,t)$) and introducing a constant force term in the
equation for $y(x,t)$. These requirements lead to the following
problem for $h_0(x,t)$
\begin{equation}
\Gamma \frac{\partial^2 h_0}{\partial x_{||}^2}=F-m_c,
\label{eq:4a}
\end{equation}
with the boundary conditions in eq. (\ref{eq:3}). The solution
of this problem is given by
\begin{equation}
h_0(x,t)=\frac{F-m_c}{2\Gamma}x_{||}^2-\left(ct+
\frac{F-m_c}{2\Gamma}L^2\right)\frac{x_{||}}{L}+ct.
\label{eq:4b}
\end{equation}
Then, substituting eq. (\ref{eq:4}) in eq. (\ref{eq:2}), with
$h_0(x,t)$ given by eq. (\ref{eq:4b}), and taking the limits
$vt\gg(F-m_c)L^2/2\Gamma$ and $x_{||}\ll L$ we obtain the
following equation for $y(x,t)$
\begin{equation}
\lambda \partial_t y=\Gamma\nabla^2y+
F-\lambda c+\eta(x,ct+y).
\label{eq:7}
\end{equation}
This equation describe the fluctuations of the interface profile
$h(x,t)$ around $h_0(x_{||},t)$, away from the boundaries and
for very long times. The constant force $F$, introduced in eq.
(\ref{eq:4a}), will be determined self-consistently using the
constraint $\langle y(x,t)\rangle=0$. Within this approximations
the anisotropy introduced by the boundary conditions, such that
the resulting equation for $y(x,t)$ is isotropic. The influence
of the anisotropy will be considered in the next subsection to
compute the avalanche exponents.

Eq. (\ref{eq:7}) is identical to the one obtained for the
fluctuations of an elastic interface driven uniformly by a
constant force $F$ \cite{nattermann,narayan1}. However, in this
case $F$ is a fixed parameter while the interface velocity $v$
($c$ in eq. (\ref{eq:7})) is obtained self-consistently from the
equation of motion. A depinning transition takes place at
certain critical force $F_c$ determined by the disorder. For
$F<F_c$ the interface is pinned after certain finite time while
for $F>F_c$ it moves with finite average velocity
$v\sim(F-F_c)^\beta$. On the contrary, in our model the
interface velocity $c$ is the fixed parameter, while $F$ is
determined self-consistently from the equation of motion. Since
$c>0$ the system will always be above the depinning transition,
i.e. $F>F_c$. Moreover, to obtain an average interface velocity
$c$ we should have $c\sim(F-F_c)^\beta$ and therefore
\begin{equation}
F=F_c+\text{const.}c^{1/\beta}.
\label{eq:8}
\end{equation}
To reach the critical state $F=F_c$ we must then fine-tune $c$
to zero. In other words, the critical state will be obtained in
the limit of separation of time scales $c\rightarrow0$, which is
usually satisfied in numerical simulations. According to eq.
(\ref{eq:4b}) adjusting the constant force $F$ we are just
adjusting the curvature of the interface profile $h(x,t)$ along
the $x_{||}$ direction. Hence, the system  self-organizes into a
stationary state where the curvature of the interface balances
the pinning forces. This conclusion, which was already
conjectured by Paczuski and Boettcher, was obtaining here by
solving the LID model dragged at one end.

\subsection{Scaling exponents}

The fluctuations around the average profile are characterized by
the pair correlation function. At the critical state the pair
correlation function is expected to obey the scaling law
\cite{nattermann,narayan1}
\begin{equation}
\langle[y(x,t)-y(0,0)]^2\rangle \sim
|x|^{2\zeta}g(t/|x|^z),
\label{eq:11}
\end{equation}
where $g(x)$ is an scaling function and $z$ and $\zeta$ are the
dynamic and roughness exponents, respectively. These exponents
are obtained through a RG analysis of eq. (\ref{eq:7})
\cite{nattermann,narayan1,narayan2}. The upper critical
dimension is found to be $d_c=4$ and below $d_c$ the scaling
exponents are given by
\begin{equation}
z=2-\frac{2}{9}(4-d),\ \ \ \ \zeta=\frac{4-d}{3},
\label{eq:12}
\end{equation}
for a random distribution of pinning forces
\cite{nattermann,narayan1} and
\begin{equation}
z=2-\frac{4-d}{3},\ \ \ \ \zeta=0,
\label{eq:12a}
\end{equation}
for periodic distribution of pinning \cite{narayan2}. The
exponents $z$ and $\zeta$ are thus identical to those obtained
for the constant force case.

However, other scaling exponents result different because of the
anisotropy introduced by the boundary conditions. In the case of
an interface driven by a constant force the average interface
profile above the depinning transition is given by $vt$ and,
therefore, there is no preference direction in space. On the
contrary, in the case of an interface dragged at one end the
average interface profile is given by $h_0(x_{||},t)$, which is
clearly non-uniform along the $x_{||}$ direction as one can see
from eq. (\ref{eq:4b}).

To analyze the influence of the anisotropy introduced by the
boundary conditions let us analyze the dynamics of active sites.
Let $\rho_a(x_{||},x_\bot,t)$ the average density of active
sites at site $(x_{||},x_\bot)$ and time $t$, given was a site
active at $x_{||}=0$. Here $x_{||}$ as above denotes the
position along the preferential direction and $x_\bot$ is a
$d-1$ dimensional vector denoting the position in a plane
perpendicular to $x_{||}$. Since the flow of grains takes place,
in average, along the positive $x_{||}$ direction then the
average flux of particles through the plain $x_{||}=L$ is given
by
\begin{equation}
J(L)=\int dt d^{d-1}x_\bot\rho_a(L,x_\bot,t).
\label{eq:12b}
\end{equation}
On the other hand, at the critical state $c\rightarrow 0$ the
average density of active sites should satisfy the scaling law
\begin{equation}
\rho_a(x_{||},x_\bot,t)=t^{\eta-d/z}f\left(\frac{x_{||}}{t^{1/z}},
\frac{|x_\bot|}{t^{1/z}},\frac{t}{L^z}\right),
\label{eq:12c}
\end{equation}
where $\eta$ is another scaling exponent and $f$ is an scaling
function. Then, substituting this expression in eq.
(\ref{eq:12b}) it results that
\begin{equation}
J(L)\sim L^{(1+\eta)z-1}.
\label{eq:13c}
\end{equation}
Now, if the system is in a stationary state for each grain we
put at $x_{||}=0$ one grain should go out at $x_{||}=L$ and,
therefore, $J(L)=1$. This stationary condition will be satisfied
only if
\begin{equation}
(1+\eta)z=1,
\label{eq:14c}
\end{equation}
in eq. (\ref{eq:13c}. A more familiar scaling relation is
obtained if one computes the mean avalanche size, which is given
by
\begin{equation}
\langle s\rangle=\int dt dx_{||}d^{d-1}x_\bot \rho_a(x_{||},x_\bot,t).
\label{eq:14d}
\end{equation}
Substituting the scaling law for $\rho_a$ in eq. (\ref{eq:12c})
in this expression it results that
\begin{equation}
\langle s\rangle \sim L^{(1+\eta)z}\sim L,
\label{eq:14e}
\end{equation} 
which is the usual scaling dependency of the mean avalanche size
with system size in critical slope models. This derivation using
such simple scaling arguments is reported here for the first
time.

Using the RG estimates for the exponents $z$ and $\zeta$ in eqs.
(\ref{eq:12}) or (\ref{eq:12a}) and the scaling relation
$\langle s\rangle\sim L$ we can compute the avalanche scaling
exponents. Let $s$ be the avalanche size and $T$ its duration,
which are distributed according to $P(s)$ and $P(T)$,
respectively. Just at the critical state one expect that these
distributions satisfy the power law behavior $P(s)\sim
s^{-\tau_s}$ and $P(T)\sim T^{-\tau_t}$, where $\tau_s$ and
$\tau_t$ are the avalanche distribution exponents. However, for
a system of finite size a characteristic avalanche size $s_c\sim
L^D$ and duration $T_c\sim L^z$ will appear, where $D$ is the
avalanche fractal dimension. Near the critical state the
distributions of avalanche size and duration will thus satisfy
the scaling laws
\begin{equation}
P(s)\sim s^{-\tau}f(s/L^D),\ \ \ \ P(T)\sim T^{-\tau_t}g(T/L^z),
\label{eq:13}
\end{equation}
where $f(x)$ and $g(x)$ are some cutoff functions with the
asymptotic behaviors $f(x),g(x)\sim 1$ for $x\ll1$ and
$f(x),g(x)\ll1$ for $x\gg1$.

The exponents $\tau_s$, $\tau_t$, D and $z$ are not all
independent. Since $s\sim T^{z/D}$ then the condition $\int
dsP(s)=\int dTP(T)$ implies
\begin{equation}
(\tau_s-1)D=(\tau_t-1)z. 
\label{eq:14}
\end{equation}
Another scaling relation can be obtained taking into account
that $\langle s\rangle\sim L$ (see eq. (\ref{eq:14e}), resulting
\begin{equation}
(2-\tau_s)D=1.
\label{eq:15}
\end{equation}
Then, from eqs. (\ref{eq:14}) and (\ref{eq:15}) it follows that
\begin{equation}
\tau_s=2-\frac{1}{D},\ \ \ \ \tau_t=1+\frac{D-1}{z}.
\label{eq:15a}
\end{equation}
Finally there is a scaling relation which relates the avalanche
dimension exponent $D$ with the roughness exponent $\zeta$.
Below the upper critical dimension the avalanches are compact
objects \cite{paczuski3} and therefore $s\sim \Delta h r^d$,
where $\Delta h$ is the characteristic fluctuation of the
interface width during the avalanche and $r$ its characteristic
linear extent in the $d$-dimensional substrate. Then since
$\Delta h\sim r^\zeta$ and $s\sim r^D$ one obtains
\cite{paczuski3}
\begin{equation}
D=d+\zeta
\label{eq:16}
\end{equation}
Above the upper critical dimension the avalanches are no more
compact and $D=d_c=4$ \cite{paczuski3}.

\section{Comparison with experiments and numerical simulations}

Using the values of $z$ and $\zeta$ obtained from the RG
analysis in eq. (\ref{eq:12}) or (\ref{eq:12a}) and the scaling
relations in eqs. (\ref{eq:15a}) and (\ref{eq:16}) we can
determine all the avalanche exponents. The results in one and
two dimensions are shown in table \ref{tab:1} and \ref{tab:2}
for random and periodic pinning, respectively. Some numerical
estimates for critical slopes model are also shown for
comparison.

In $d=1$ we count with numerical simulations of ricepile models,
which are critical slope sandpile models with certain randomness
in the toppling rule. For instance Frette \cite{frette}
considered a ricepile model where the slope threshold $m_c$ is
selected at random after each toppling. A modified version of
this model was later used by Christensen {\em et al}
\cite{christensen}. In a somehow different model Nunes-Amaral
and Lauritsen \cite{nunes1} considered a rice pile model where
the toppling rule is stochastic in certain range of slopes
$m_{c1}<m<m_{c2}$ while it is deterministic above $m_{c2}$. All
these ricepile models give the same avalanche exponents and
should belong to the random disorder universality class. In
table \ref{tab:1} we display the more accurate numerical
estimates reported by Nunes-Amaral and Lauritsen.

In $d=2$ we count with a critical slope model introduced by
Bassler and Paczuski \cite{bassler} to describe the avalanche
dynamics in superconducting vortexpiles. In their model $z_i$ is
the density of vortex at site $i$. The evolution rules are not
exactly those described above but a critical slope condition was
used. They also considered quenched random pinning forces and
therefore it should belong to the random disorder universality
class.

More recently Cruz, Mulet and Atshuler \cite{cruz} have made
simulations of the model introduced by Bassler and Paczusky but
considering both random and periodic pinning forces. In the case
of random pinning their numerical estimates reproduced, within
the numerical error, those reported by Bassler and Paczuski
\cite{bassler}. However, in the case of periodic pinning they
obtained very different exponents (see table \ref{tab:2})
suggesting that there are two universality class. Within our
analysis the existence of these two universality classes is
clear, corresponding with random and periodic distributions of
pinning forces.

In all cases we observe a very good agreement between the
numerical estimates and our predictions. This agreement is more
surprising because the dimensions considered are far from the
upper critical dimension $d_c=4$. In table \ref{tab:1} we also
display the scaling exponents $D$ and $z$ obtained from
numerical simulations of the LID model driven at constant force,
which are expected to be identical to those for the LID model
driven at one end. As one can see the agreement is quite good,
event better than with the RG estimates.

Recently Tadi\'c and Nowak \cite{tadic} has observed that the
scaling exponents of the random-field Ising model (RFIM) in the
low disorder regime are very close to those of the ricepile
model (see table \ref{tab:1}). In particular they considered a
diluted two-dimensional Ising model with weak random fields and
with an anisotropic initial condition. In the low disorder
regime a single domain wall separating two regions with
different spin orientations is observed. In this regime the
Barkhausen avalanches are attributed to the fluctuations in the
domain wall. The agreement with the exponents for the ricepile
model suggests that the domain walls in this regime may be also
described by the equation of motion \ref{eq:2}, while the
anisotropy was introduced through the initial conditions. On the
contrary, on the high disorder regime a multi-domain structure
is obtained, resulting in different avalanche exponents.

In this analysis we have not included a comparison with some
available experimental results for sandpiles \cite{held},
ricepiles \cite{frette1} and vortexpiles \cite{superconductors}
because it is very difficult to compare our results with
experimental measurements. In most of the experiments the
measured magnitude is the number of grains (vortexes) leaving
(entering) the system which will be denoted by $s_0$. In general
$s_0$ can be also described by the scaling law $P(s_0)\sim
s_0^{-\tau_0}f(s_0/L^{D_0})$. However, in average, for one grain
(vortex) entering the system one grain (vortex) goes out and
therefore $\langle s_0\rangle=1$. Then if one computes $\langle
s_0\rangle$ using the scaling law for $P(s_0)$ one obtains
$\langle s_0\rangle\sim L^{(2-\tau_0)D_0}\sim1$, which
immediately gives $\tau_0=2$. This value is near the one
reported for real sandpiles, ricepiles and vortex piles. On the
contrary, in numerical simulations and in the model analyzed in
this work the avalanche size $s$ is given by the number of
toppling events which is not necessarily proportional to the
flow of grains (vortexes), the magnitude measured in
experiments. In fact, in the case of the avalanche size $s$, the
conservation law yields $\tau_s=2-1/D<2$) (see eq.
(\ref{eq:15}). Hence, we conclude that $\tau_0\neq\tau_s$.

\section{conclusions}

We conclude that ricepile models, the vortexpile model by
Bassler and Paczuski, and the RFIM in the low disorder regime
belong to the same universality class, that of critical slope
sandpile models. The prototype model for this universality class
is the LID model dragged at one end, as it was already
conjectured by Paczuski and Boettcher \cite{paczuski}. Our
analytical treatment has demonstrated that their guess was
correct. For the first time, we have solved the LID model
dragged at one end, obtaining the complete set of scaling
exponents. Some of these results were supported by previous RG
calculations developed for constant force LID but others, like
the linear dependence of the susceptibility with system size,
are intrinsic to this model. Moreover, as in the constant force
case, we found two different universality classes corresponding
with random and periodic pinning forces. Our predictions were
found in very good agreement with numerical simulations of
different critical slope models.

\section*{ackowledgements}

This work has been partly supported by the {\em Alma Mater }
prize, from the Havana University.


\begin{table}\narrowtext
\begin{tabular}{lllllll}
$d$ & Model & $\tau$ & $\tau_t$ & $D$ & $z$ & Ref.\\ \hline
1 & RP & 1.53(5) & 1.84(5) & 2.20(5) & 1.40(5) & \cite{nunes1}\\
  & RFIM & 1.58 & 1.89 & 2.23 & 1.45 & \cite{tadic}\\ 
  & LID & & & 2.25 & 1.42(3) & \cite{leschhorn}\\
  & RG & $\frac{3}{2}=$1.5 & $\frac{7}{4}=$1.75 & 2 & 
  $\frac{4}{3}\approx$1.33\\
\\
2 & VP & 1.63(2) & 2.13(14) & 2.7(1) & 1.5(1) & \cite{cruz}\\
  & LID & & & 2.75(2) & 1.58(4) & \cite{leschhorn}\\
  & RG & $\frac{13}{8}=$1.625 & $\frac{29}{14}\approx$2.07 & 
  $\frac{8}{3}\approx$2.67 & $\frac{14}{9}\approx$1.56
\end{tabular}
\caption{Avalanche scaling exponents of the ricepile model (RP),
the vortexpile model (VP), the random-field Ising model (RFIM),
and the LID model (LID), together with our estimates using
previous RG calculations. Error bars are indicated between
parenthesis.}
\label{tab:1}
\end{table}

\begin{table}\narrowtext
\begin{tabular}{lllllll}
$d$ & Model & $\tau$ & $\tau_t$ & $D$ & $z$ & Ref.\\ \hline
1 & RG & 1 & 1 & 1 & 1\\
\\
2 & VP & 1.45(2) & 1.70(8) & 2.2(1) & 1.6(1) & \cite{bassler}\\
  & RG & $\frac{3}{2}=$1.5 & $\frac{7}{4}=$1.75 & 2 & 
  $\frac{4}{3}\approx$1.33
\end{tabular}
\caption{Avalanche scaling exponents of the vortexpile (VP)
model with periodic pinning, together with our estimates using
previous RG calculations. Error bars are indicated between
parenthesis.}
\label{tab:2}
\end{table}

\end{multicols}


\begin{thebibliography}{50}

\bibitem{richter} B. Gutenberg and C. F. Richter, Ann. Geophys.
{\bf 9}, 1(1956).

\bibitem{held} G. A. Held, D. H. Solina, D. T. Keane, W. J.Haag,
and G.  Grinstein, Phys. Rev. Lett. {\bf 65}, 1120 (1990).

\bibitem{frette1} V. Frette, K. Christensen, A.
Malthe-Sorenssen, J. Feder, T. Tossang, and P. Meakin, Nature
{\bf 3}, 22 (1996).

\bibitem{superconductors} S. Field, J. Witt, F. Nori, and X.
Ling, Phys. Rev. Lett. {\bf 74}, 1206 (1995); K. Behnia and C.
Capan, cond-mat/9902334.

\bibitem{barkhausen} B. Alessandro, C. Beatrice, G. bertotti,
and A. Montorsi, J. Appl. Phys. {\bf 68}, 2908 (1990); J. S.
Urbach, R. C. Madison, and J. T. Markert, Phys. Rev. Lett. {\bf
75}, 276 (1995); \bibitem{durin1} G. Durin, G. Bertotti, and A.
Magni, Fractals {\bf 3}, 351 (1995).

\bibitem{crack} A. Petri, G. Paparo, A. Vespignani, A. Alippi,
and M. Constantini, Phys. Rev. Lett. {\bf 73}, 3423 (1994).

\bibitem{bak2} P. Bak, C. Tang, and K. Wiesenfeld, Phys. Rev.
Lett. {\bf 59}, 381 (1987); Phys. Rev. A {\bf 38}, 364 (1988).

\bibitem{kadanoff} L. P. Kadanoff, S. R. Nagel, L. Wu, and S.
Zhou,Phys. Rev A {\bf 39}, 6524 (1989).

\bibitem{manna1} S. S. Manna, J. Phys. A {\bf 24}, L363 (1991).

\bibitem{frette} V. Frette, Phys. Rev. Lett. {\bf 70}, 2762 (1993). 

\bibitem{nunes1} L. A. Nunes-Amaral and K. B. Lauritsen, Phys.
Rev. E. {\bf 54}, R4512 (1996).

\bibitem{zhang} Y.-C. Zhang, Phys. Rev. Lett. {\bf 63}, 470 (1989).

\bibitem{olami} Z. Olami, H. J. S. Feder, and K. Christensen,
Phys. Rev. Lett. {\bf 68}, 1244 (1992).

\bibitem{bassler} K. E. Bassler and M. Paczuski, Phys. Rev.
Lett. {\bf 81}, 3761 (1998).

\bibitem{christensen} K. Christensen, \'A. Corral, V. Frette, J.
Feder, and T. Jossang, Phys. Rev. Lett. {\bf 77}, 107 (1996).

\bibitem{manna2} S. S. Manna, Physica A {\bf 179}, 249 (1991).

\bibitem{MF} C. Tang and P. Bak, J. Stat. Phys. {\bf 51}, 797
(1988); A. Vespignani and S. Zapperi, Phys. Rev. Lett. {\bf 78},
4793 (1997).

\bibitem{vespignani1} A. Vespignani, R. Dickman, M. A. Mu\~noz
and S. Zapperi, cond-mat/9806249.

\bibitem{diaz-guilera} A. D\'{\i}az-Guilera, Europhys. Lett. {\bf
26}, 177 (1994).

\bibitem{pietronero} L. Pietronero, A. Vespignani, and S.
Zapperi, Phys. Rev. Lett. {\bf 72}, 1690 (1994).

\bibitem{paczuski} M. Paczuski and S. Boettcher. Phys. Rev.
Lett. {\bf 77}, 111 (1996).

\bibitem{nattermann} T. Nattermann, S. Stepanow, L.-H. Tang, and  H.
Leschhorn, J. Phys. II France {\bf 2},1483 (1992).

\bibitem{narayan1} O. Narayan and D. S. Fisher, Phys. Rev. B
{\bf 48}, 7030 (1993).

\bibitem{narayan2} O. Narayan and D. S. Fisher, Phys. Rev. Lett.
{\bf 68}, 3615 (1992).

\bibitem{leschhorn} H. Leschhorn, Physica A {\bf 195}, 324 (1993). 

\bibitem{paczuski3} M. Paczuski and S. Boettcher, Phys. Rev. E
{\bf 53}, 414 (1996).

\bibitem{cruz} R. Cruz, R. Mulet, and E. Atshuler, Physica A
(submitted).

\bibitem{tadic} B. Tadi\'c and U. Nowak, cond-mat/9903090; B.
Tadi\'c, preprint.

\end{thebibliography}
\end{document}